\documentstyle[osa,manuscript]{revtex}  

\begin{document}

\title{%
Micro-Raman scattering investigation of MgB$_2$ and RB$_2$ (R=Al, Mn, Nb and Ti)}
\author{N.~Ogita$^{1}$, T.~Kariya$^{2}$, K.~Hiraoka$^{2}$, J.~Nagamatsu$^{3}$, 
T.~Muranaka$^{3}$, H.~Takagiwa$^{3}$, J.~Akimitsu$^{3}$ and M.~Udagawa$^{1}$}
\address{%
$^{1}$Faculty of Integrated Arts and Sciences, Hiroshima University, Hiroshima 739-8521, Japan\\
$^{2}$Faculty of Engineering, Ehime University, Matsuyama 790-8577, Japan\\
$^{3}$Department of Physics, Aoyama-Gakuin University, Tokyo 157-8521, Japan\\
}%

\maketitle
\begin{abstract}
Phonon spectra have been investigated by micro-Raman scattering from 290 to 20K. 
New peaks at 500 and 674 cm$^{-1}$ appear below 250K in MgB$_2$. 
These peaks are explained by the Fermi resonance, 
where Raman-active E$_{2g}$ intereacts with the overtone of E$_{1u}$ due to anhamonicity. 
Raman spectra of the isostructural RB$_2$ (R=Al, Mn, Nb and Ti) have been also measured 
at 290K. The line width of E$_{2g}$ phonon of the superconductor 
MgB$_2$ and NbB$_2$ shows the broader than that of the non-superconductors. 
It is found that the anharmonicity of phonons is important for the superconductivity 
for MgB$_2$.

\end{abstract}

\pacs{78.30.-j, 74.25.Kc, 63.20.Kr, 63.20.Ry}
\vspace{2em}

Soon after the discovery of high-T$_C$ superconductivity of MgB$_2$ by Akimitsu and 
his coworkers~\cite{1}, 
it was revealed by the isotope effect by the substitution of $^{11}$B~\cite{2} 
and the observation of a coherent peak in the NMR experiment~\cite{3} 
that MgB$_2$ is a BCS-type supercondoctor with the electron-phonon coupling 
and its symmetry is a s-wave. 
With increasing the applied pressure, the transition temperature (T$_C$)
decreases\cite{4,5}, but, this pressure dependence is not 
explained by the simple BCS theory. 
Loa and Syassen~\cite{6} explained the pressure dependence of T$_C$ by the combination 
of increasing phonon frequency and decreasing electronic density of states at Fermi surface.
These feature of MgB$_2$ is related to the layered structure, 
consisting of the alternating stacking of the hexagonal layer of Mg 
and the graphite-like layer of B along the $c$-axis. 
Some reports emphasize that the layered B state plays the most important 
role of the appearance of superconductivity.\cite{7,8,9} 
Thus, the microscopic origin of high-T$_C$ superconductivity in MgB$_2$ 
is not fully understood and the important knowledge may be the atomic 
interaction of inter- and intra-plain. 

Raman spectroscopy is an excellent technique to investigate elementary excitation, 
such as phonons. The P6/mmm symmetry of the MgB$_2$ structure gives us 
the following phonon number for each irreducible representations at 
Brillouion zone center; $\Gamma = 2\mbox{A}_{2u} + \mbox{B}_{1g} + \mbox{E}_{2g} 
+ 2\mbox{E}_{1u}$. 
The Raman active phonon is only E$_{2g}$, which is the out-of-phase vibration of the 
adjacent B atoms in the $c$-plane. In the Raman scattering, 
the phonon due to Mg cannot be observed, since Mg locates just on the inversion 
center.

The Raman scattering spectra of MgB$_2$ have been reported by several group
\cite{10,11,12}. Bohnen {\it et al.} assigned the broad mode at 
580 cm$^{-1}$ to be the E$_{2g}$ mode from the agreement of phonon energies with 
their lattice dynamical calculation~\cite{10}. 
In general, for this crystal structure, the E$_{2g}$ mode should be the highest 
energy phonon than that of the B$_{1g}$ mode, which is the out of plane mode of B. 
However, in the actual MgB$_2$, the energy of E$_{2g}$ mode is lower than that of 
B$_{1g}$ due to the strong electron-phonon coupling~\cite{10,13}. 
Alexander {\it et al.} also concluded that the E$_{2g}$ phonon is strongly 
anharmonic and couples to electronic excitations~\cite{11}. 
Martinho {\it et al.}, however, 
concluded from the resonant Raman experiment that the E$_{2g}$ phonon near 
the $\Gamma$ point may not play an important role in the 
electron-phonon coupling for the superconductivity in MgB$_2$~\cite{12}. 
Although MgB$_2$ is the superconductor due to the phonon mechanism, 
the role of the phonons is still unclear, even the E$_{2g}$ phonon. 

In this letter we present the Raman scattering spectra measured by the microscope system, 
to obtain the pure spectra from the superconducting part, 
since the sample was pollycrystalline. We also studied the other isostructural 
crystals for the comparison with MgB$_2$.

For MgB$_2$, we employed three different samples; the commercial one (Rare metallic Co. Japan), 
the prepared sample by a HIP, and the sample annealed in a high pressure after 
the HIP preparation. The Raman scattering spectra shown in this report are the spectra from 
the last annealed sample. Other samples of RB$_2$ (R=Al, Mn, Nb, and Ti) were grown by 
a sintering. 

In Raman scattering experiments, the laser beam of 514.5 and 488 nm from an Ar ion laser 
was employed as an incident light. 
In the micro-Raman scattering system attatched with the $\times$50 objective, 
the incident light was focused in a $\sim$2$\mu$m area in diameter on the sample surface. 
While, the size of the focused area became $\sim$80$\mu$m in diameter in the macro-Raman 
scattering experiments. 
The scattered light was analyzed by a triple-monochrometer~(JASCO, NR-1800) and detected 
by a multichannel CCD detector cooled by a lig.~N$_2$~(Princeton Instruments). 
In the measurements of the low temperatures, we employed a $^4$He-flow type cryostat for 
both macro- and micro-Raman scattering. 
To avoid local heating by the incident light, a $^4$He heat-exchange gas was filled in 
the sample cell. 
The polarization depndence was not measured, since the sample was polycrystalline. 

Figure 1 is the photograph of the polished surface of the annealed sample of 
MgB$_2$ after the HIP preparation. 
As shown in Fig.1, three different color grains of gold, blue and black are recognized. 
The grain  size are about 3$\sim$5 $\mu$m. 
The dark blue spot at the center of the photograph corresponds to the incident light. 
The Raman scattering spectra from each grain are shown in Fig.2. 
Comparing these spectra with our macro-Raman results and 
reported ones~\cite{10, 11, 12}, 
the gold grain is regarded as MgB$_2$. The additonal peaks observed in other grains are 
the spectra from the following impurities. 
The existence of a small amount of MgB$_4$, MgB$_6$, MgB$_{12}$, MgO, and MgCO$_3$ 
has been observed by the X-ray diffraction.

Figure 3 shows the temperature dependence of the micro-Raman scattering 
spectra from the gold grain measured with the $\times$ 50 objective. 
Each spectrum is shifted along the vertical axis to avoid crossing. 
The peaks, labelled by astarisks, are phonons from the blue or black grain.
As shown in Fig.~3, new peaks labelled by arrows appear at 500 and 
674 cm$^{-1}$ below T$\simeq$250K. 
Their energy, line width, and the relative intensity 
I(500cm$^{-1}$)/I(674cm$^{-1}$) do not depend on temperature. 
These peaks are originated from the Raman process, since we have obtained 
the same spectra in the different excitation energy. 
We note that the intensity of the broad spectra from 400 to 800 cm$^{-1}$ 
observed at 290K does not grow at the low temperatures. 

Figure 4 shows the temperature dependence of the macro-Raman scattering 
spectra. 
As shown in Fig.~4, only E$_{2g}$ phonon is observed for all temperatures. 
At 290K, the spectra are similar with those measured by micro-Raman 
scattering, because the gold region is dominant on the sample surface. 
In this macro-Raman spectra, the intensity of $E_{2g}$ increases 
at the lower temperatures.
In addition, the new peaks have not been observed. 
Therefore, the spectra of the macro-Raman spectra become different from 
the micro-one below T$\simeq$250K. 

To check this spactral difference between the micro- and macro-Raman 
scattering in T$<$250K, we changed the magnification ratio of the objectives. 
New peaks cannot be observed for the larger spot size of the incident beam 
than the grain size of 5$\mu$m. 
We have not observed the new peaks for the commercial and the HIP-prepared 
samples, where the grain size of the gold part is much less than 1$\mu$m. 
Thus, the grain size at the focused area is important to oberved the new 
peaks. 

The inset of Fig.~4 shows the temperature dependence of the energy and 
line width of the E$_{2g}$ phonon. 
The line width is shown by vertical solid bar. 
In the temperaure region above T$\simeq$250K, 
the peak energy increases with decreasing temperature, 
while below T$\simeq$250K the energy is almost constant. 
The solid line shows the estimated phonon energy from the reported 
Gr$\ddot{\mbox{u}}$neissen parameter
($\gamma$=$\mbox{dln}\nu /3\mbox{dln}a$=3.9)~\cite{11} 
and lattice constants~\cite{14}. 
As shown in the inset of Fig.~4, the temperature dependence of the E$_{2g}$ 
phonon energy is explained by the ordinary thermal expansion for T$<$250K, 
but the deviation of the observed energy from the estimated energy is 
clearly seen for T$>$250K. 

Here we summarize the experimental results of the temperature dependence 
of MgB$_2$. 
We have found the characterstic temperature T$\simeq$250K in MgB$_2$, where the
new peaks appear in micro-measurement and also the temperture dependence of 
E$_{2g}$ phonon energy deviates from the expected energy of the thermal expansion. 
Furthermore, the appearance of the new peaks is related to phonon coherence, 
because this depends on the spacial length ratio between the grain size and the 
laser spot size. 

Next, we discuss about the new peaks at 500 and 674 cm$^{-1}$ observed 
in micro-measurement for T$<$250K. 
The two peaks cannot be assigned as the first order phonon, because of 
no structural change. 
As shown in Fig.~4, the E$_{2g}$ phonon intensity becomes large with 
decreasing tempreature in the macro-measurement. 
If we apply this temperature dependence of the macro-one to the micro-case, 
the spectrum should be observed at low temperature, as shown by (a) in Fig.~3. 
However, only two peaks are observed in the actual spectra without the 
increase of the back ground intensity. 
This fact suggests that the intensity of E$_{2g}$ transfers to the new peaks 
below T$\simeq$250K. 
In addition, the E$_{2g}$ phonon is observed at nearly center of the two peaks 
as shown in Fig.~3. 
Therefore, these intensity transfer and energy distribution show the existence of 
the coupling between E$_{2g}$ and the others.

The coupled modes with E$_{2g}$ should be the well defind mode, since the 
line width of the new peaks is too narrow. From this point of view, 
the electronic case is excluded, since its line width might be broad. 
Thus, the possible excitation in this crystal is phonon.

The coupling mode should satisfy the following conditions; 
(1){\it the same symmetry 
respresentation} of E$_{2g}$ and (2){\it the close energy to} E$_{2g}$ 
{\it phonon}. 
In the first order process, we cannot find such phonons. 
Then, the plausible mode is the overtone of E$_{1u}$, because the 
E$_{1u}$ mode is also in-plain vibration of B as same as E$_{2g}$ mode. 
In fact, the overtone modes of E$_{1u}$ includes the symmetry of E$_{2g}$ from 
$\mbox{E}_{1u}\times\mbox{E}_{1u}=\mbox{A}_{1g}+\mbox{E}_{1u}+\mbox{E}_{2g}$ 
and its twice of energy is very close to that of the E$_{2g}$ phonon, 
because the calculated energies of E$_{1u}$ phonon are 250-300 cm$^{-1}$ and 
322cm$^{-1}$ at Brillouin zone boundary and center~\cite{10}, 
respectively. 
As the coupling between the overtone mode and one phonon, Fermi resonance 
is known~\cite{15}. 

The detailed procedure of the Fermi resonance is represented in Ref.~\cite{16}.
The Fermi coupling constant (K), bare phonon energy of E$_{2g}$ 
($\Omega_1$) and that of the overtone E$_{1u}$ (2$\Omega_2$) 
are described by 
\begin{eqnarray}
\mbox{K} &=& \sqrt{2}[\Omega_+ - \Omega_-]
\frac{(1-\mbox{R}\mbox{R}_0)(\mbox{R} + \mbox{R}_0)}
{(1-\mbox{R}\mbox{R}_0)^2 + (\mbox{R} + \mbox{R}_0)^2}\nonumber\\
\Omega_1 &=& \frac{1}{2}[\Omega_+ + \Omega_-] 
- \frac{1}{2}\sqrt{(\Omega_+ - \Omega_-)^2 - 2K^2} \nonumber\\
2\Omega_2 &=& \frac{1}{2}[\Omega_+ + \Omega_-] 
+ \frac{1}{2}\sqrt{(\Omega_+ - \Omega_-)^2 - 2K^2}, \nonumber
\end{eqnarray}
where R and R$_0$ are the ratio of the Raman polarizability matrix elements 
for the resonated case and the uncoupled one~(R=$\alpha_{0-}/\alpha_{0+}$ 
and R$_0$=$\alpha_{02}/\alpha_{01}$), respectively. 
$\Omega_+$ and $\Omega_-$ are the frequencies in the resonance. 
In the above equatioins, the observed parameters are $\Omega_+$ and $\Omega_-$, 
and R$^2$ which corresponds to the intensity ratio I($\Omega_-$)/I($\Omega_+$). 
While the unkwon parameters are R$_0$, K, $\Omega_1$ and $\Omega_2$. 
In this analysis, we assume $\Omega_1$=620cm$^{-1}$ because the $E_{2g}$ mode 
observed by the macro-mesurements does not show any coupling in all 
temperatures as shown in Fig.4.
Then, using the observed values of $\Omega_+$=674 cm$^{-1}$, $\Omega_-$=500 cm$^{-1}$, 
R=2.41 and $\Omega_1$=620 cm$^{-1}$, we obtain the Fermi coupling constant 
K$\simeq$114 and bare overtone frequency $2\Omega_2$=554cm$^{-1}$ for both 
two solutions of R$_0$=1.51 and 5. 
The energy agrees with the reported value of E$_{1u}$ phonon at Brillouin zone 
boundary~\cite{10}. 

As the extension of the above discussion, we comment on the origin of 
the high temperature superconductivity in MgB$_2$. 
The mediated phonons in the mechanism of superconductivity are probably 
the E$_{1u}$ phonon, because the E$_{1u}$ phonons include the in-plain motion 
of borons. 
When the E$_{1u}$ phonons couple with E$_{2g}$ phonons, 
an effective phonon frequency increases. As the result, the T$_C$ increases. 
This consideration is not discrepant with the results 
by Martinho {\it et al.}, where the E$_{2g}$ phonon did not show the 
expected spectral change at T$_C$, 
because the E$_{2g}$ phonon plays indirectly to the superconducting mechanism. 

Finally, we present the results for the reference materials of 
RB$_2$ (R=Al, Mn, Nb, Ti and Y), where the Raman-active phonon is 
only E$_{2g}$. 
Among the materials, MgB$_{2}$ and NbB$_{2}$ (T$_C$=5K) shows 
superconductivity. 
Figure 5 shows the plot of the phonon energy and line width as a function 
of the ratio of lattice constants $(c/a)$. 
The line width is presented by the vartical bar. 
As shown in Fig.~5, the E$_{2g}$ phonon frequency decreases with increasing 
ratio of $c/a$. 
Most pronounced property is the line width, since that of the superconducting 
crystals of MgB$_{2}$ and NbB$_2$ has the broader line width, compared with 
the others. 
This suggests that the large anharmonicity of phonons is necessary for the 
appearence of superconductivity in RB$_2$ structure. 
Furthermore, the increase of $c/a$ corresponds to the increase of the 
two-dimensional character and the decrease of the inter-layer interasction. 
It can be seen from the figure that the ratio seems to be proportional 
to T$_C$. 
Thus, to obtain the higher-T$_C$ crystals, we should seek the crystals with 
the lareger c/a and the weaker interlayer-interaction between the cation R 
and B. In addtion, there remains the problem, whether such two dimensional 
character in MgB$_2$ is similar to that of the oxide superconductors of 
Sr$_2$RuO$_4$~\cite{17} or not.

This work is supported in part by a Grant-in-Aid for COE Research 
(No.13CE2002) of the Ministry of Education, Culture, Sports, Science and 
Technology of Japan. The low temperature experiments was also supported by 
the cryogenic center of Hiroshima University. 

%
\end{document}